\documentclass[preprint,showpacs,preprintnumbers,amsmath,amssymb,aps]{revtex4}

\usepackage{graphicx}
\usepackage{amssymb}
\usepackage{amsmath}
\usepackage{bm}

\usepackage{color}

\begin{document}
\setcounter{page}{0}
\title[]{Andreev-reflection spectroscopy with superconducting indium - a case study}
\author{Kurt \surname{Gloos}}
\email{kgloos@utu.fi}
\thanks{Fax: +382-2333-5470}

\affiliation{Wihuri Physical Laboratory, Department of Physics   and Astronomy, University of Turku, FIN-20014 Turku, Finland}
\affiliation{Turku University Centre for Materials and Surfaces (MatSurf), FIN-20014 Turku, Finland}

\author{Elina \surname{Tuuli}}
\affiliation{Wihuri Physical Laboratory, Department of Physics and Astronomy, University of Turku, FIN-20014 Turku, Finland}
\affiliation{The National Doctoral Programme in Nanoscience (NGS-NANO), FIN-40014 University of Jyv\"askyl\"a, Finland}

\date{\today}

\begin{abstract}
We have investigated Andreev reflection at interfaces between superconducting indium ($T_c = 3.4\,$K) and several normal conducting non-magnetic metals (palladium, platinum, and silver) down to $T = 0.1\,$K as well as zinc ($T_c = 0.87\,$K) in its normal state at $T = 2.5\,$K. 
We analysed the point-contact spectra with the modified on-dimensional BTK theory valid for ballistic transport.
It includes Dynes' quasi-particle lifetime as fitting parameter $\Gamma$ in addition to superconducting energy gap $2\Delta$ and strength $Z$ of the interface barrier. 
For contact areas from less than $1\,$nm$^2$ to $10000\,$nm$^2$ the BTK $Z$-parameter was close to 0.5, corresponding to transmission coefficients of about $80\%$, independent of the normal metal. 
The $Z$-parameter varies by less than $\pm 0.1$ around its average value, indicating that the interfaces have a negligible dielectric tunneling barrier. 
Also Fermi surface mismatch does not account for the observed $Z$.
The extracted value $Z \approx 0.5$ can be explained by assuming that practically all of our point contacts are in the diffusive regime.

\end{abstract}

\pacs{85.30.Hi, 73.40.-c, 74.45.+c}


\keywords{point contacts, metal interfaces, normal reflection, Andreev reflection}

\maketitle

\section{Introduction}

An interface between two conductors reduces charge (electron or hole) transport, transmitting a fraction $\tau$ of the incident current and reflecting the remainder $1-\tau$.
Normal reflection plays a central role in Andreev-reflection spectroscopy because also the Andreev-reflected holes can be normal reflected.
The BTK theory for ballistic transport \cite{Blonder1982} assumes that normal reflection affects them both in the same way.
This enables to measure the transmission coefficient of normal-superconductor interfaces.
Ballistic transport requires that the electron mean free path is much larger than the contact diameter $d$. 
Therefore one can reach the ballistic regime by making the contacts small enough.
When the contacts are made larger, they become diffusive. 
In that case the elastic electron mean free path $l_0$ is much smaller than the contact diameter while the inelastic one $l_{in}$ is so large that the diffusive length $\Lambda = \sqrt{l_{in} l_0 / 3}$ still exceeds the contact diameter \cite{Lysykh1980}.

Blonder and Tinkham \cite{Blonder1983} explained the Andreev reflection double-minimum structure of ballistic contacts - an enhanced resistance around zero bias inside the energy gap - with a combination of tunnelling through a dielectric layer and the mismatch of Fermi velocities. 
By approximating the real dielectric barrier of width $w$ and height $\Phi$ with a $\delta$-function of strength $Z_b = \Phi w / \hbar \sqrt{v_{F1} v_{F2}}$ and assuming free electrons with $r = v_{F1}/v_{F2}$ being the ratio of Fermi velocities $v_{F1}$ and $v_{F2}$ of the two electrodes, the transmission coefficient $\tau = 1 / (1+Z^2)$ can be obtained from \cite{Blonder1983}
\begin{equation}
  Z^2 = Z_b^2 + (1-r)^2/(4 r)
  \label{Z parameter}
  \end{equation}
Thus one could directly measure Fermi-velocity ratios once the contribution $Z_b$ of the dielectric barrier is known.

In a typical Andreev-reflection experiment a dielectric oxide \cite{Simmons1964} or water/ice layer \cite{Repphun1995} has to be expected when the two electrodes and their contact are not prepared at ultra-high vacuum.
And a junction between metals with different Fermi surfaces causes normal reflection since the electron wave functions have to be adjusted across the interface.
Even a junction between two identical metals disrupts the crystal lattice  and should lead to some amount of normal reflection. 
However, those effects are difficult to quantify. 

Further complications arose when Steglich {\it et al.} \cite{Steglich1979} discovered  heavy-fermion superconductors where the 'heavy' conduction electrons with an extremely small Fermi velocity form the Cooper pairs.
The first point-contact study of such compounds by U.~Poppe \cite{Poppe1985} and Steglich {\it et al.} \cite{Steglich1985} focussed on Giaever-type tunneling to measure the density of states of the new superconductors and the Josephson effect to probe the symmetry of the heavy-fermion order parameter, without considering Andreev reflection.
E.~W.~Fenton \cite{Fenton1985} predicted a huge normal reflection coefficient, corresponding to $Z \gg 1$, for interfaces between a conventional and a heavy-fermion metal because of Fermi velocity mismatch.
This idea got partial support by a large background residual resistance of heavy-fermion contacts where the cross-sectional area could be determined independently \cite{Gloos1995,Gloos1996a,Gloos1996b}. 
However, the expected tunneling-like Andreev reflection anomalies have not been found.

Deutscher and Nozi{\`e}res \cite{Deutscher1994} explained the weak normal reflection observed in Andreev-reflection experiments with heavy-fermion compounds by noting that the bare electrons and holes cross the interface, and not the heavy particles. 
This suggests that it is not the mismatch of Fermi velocities but that of the Fermi wave numbers that matters.
Equation (\ref{Z parameter}) remains valid with $r$ replaced by the ratio of Fermi wave numbers $k_{F1}$ and $k_{F2}$ of the electrodes.
For interfaces between heavy-fermion compounds and conventional metals this ratio is of order unity, and therefore the $Z$ parameter should be rather small.
A number of point-contact Andreev reflection experiments on heavy-fermion compounds, for example \cite{Goll1993,DeWilde1994,Naidyuk1996,Park2008}, support this interpretation. 
Because heavy fermion metals often have an intrinsically short electron mean free path, it is possible that contacts with them are not ballistic but in the diffusive limit \cite{Gloos1996c,Gloos1996b,Gloos1998}.

Also the proximity effect at superconducting - normal metal thin film layered structures depends strongly on the transparency of the interfaces \cite{Attanasio2006,Kushnir2009}.
These experiments reveal that $\tau \lesssim 0.5$ (corresponding to $Z \gtrsim 1$) for contacts between simple metals, considerably less than the expected $\tau \approx 1$ ($Z \approx 0$) in free-electron approximation.
The thin films are deposited in ultra-high vacuum, which excludes a dielectric interface barrier and leaves Fermi surface mismatch or a lattice discontinuity to explain the strong normal reflection.

One can also measure directly the current perpendicular to plane (CPP) resistance of an interface with a well-defined geometry and large cross-sectional areas of order $1\,\mu$m$^2$ \cite{Pratt2009,Sharma2009} and compare it with electronic-structure calculations \cite{Xu2006,Xu2006b}. 
The CPP resistance should contain information about normal reflection, but it is difficult to extract
because of the lacking knowledge of the resistance without normal reflection.

Measuring electron spin polarization using Andreev-reflection spectroscopy \cite{Soulen1998} is another research topic that relies heavily on normal reflection and the ballistic nature of the contacts. 
According to the generally accepted view \cite{Bugoslavsky2005,Baltz2009}, the true spin polarization is obtained at highly transparent interfaces when $Z \rightarrow 0$ while the measured polarization drops with increasing normal reflection.
This strong $Z$ dependence of the polarisation does not match the results of the Tedrow-Meservey tunneling experiments \cite{Tedrow1973} performed in the opposite $Z \gg 1$ limit, possibly indicating that the interface transparency affects the measured polarization in a complicated way \cite{Kant2002,Woods2004}.

\begin{figure}
\includegraphics[width=12cm]{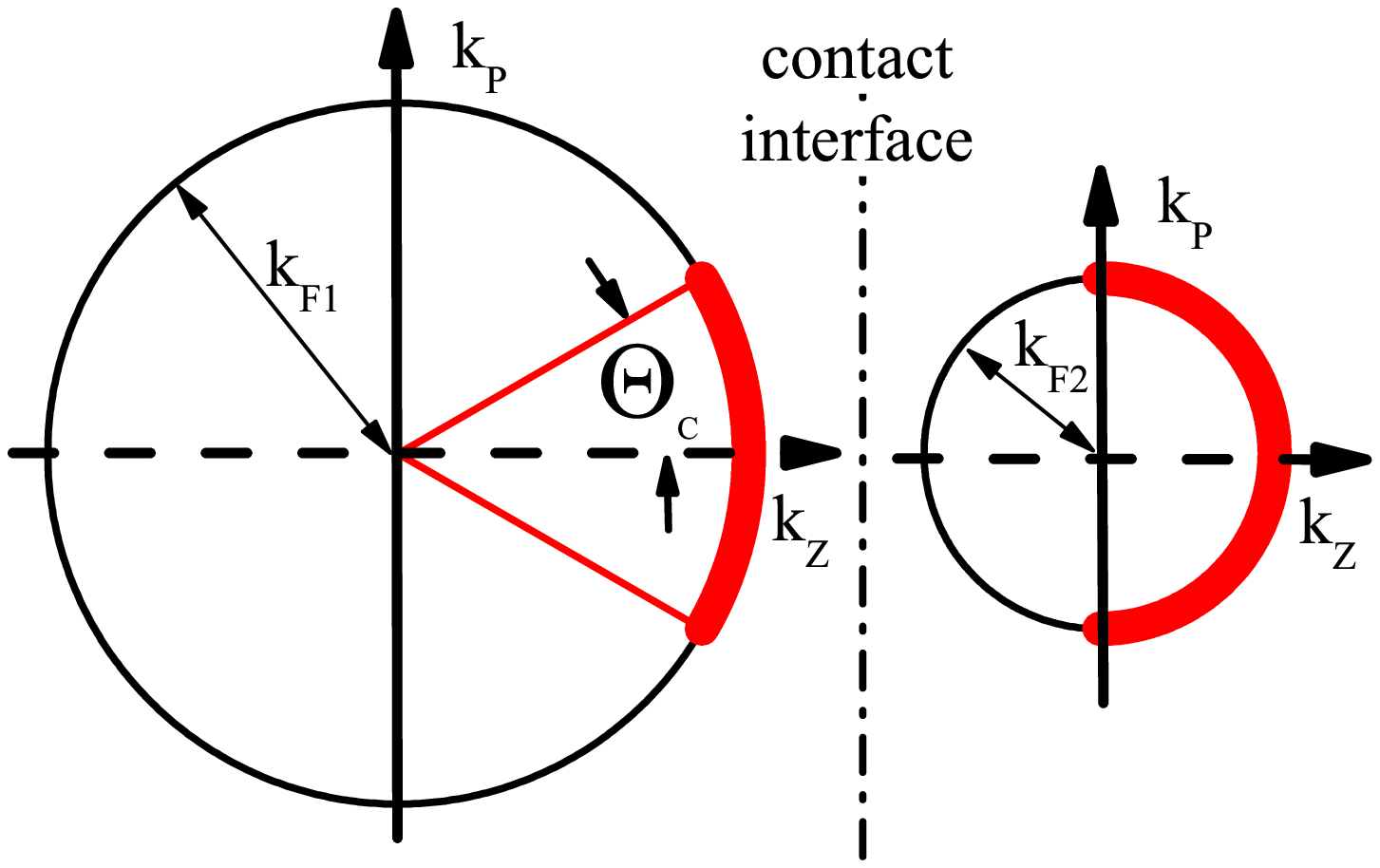}
\caption{(Color online) Schematics of a point contact between two different metals in momentum space with $k_{F1} > k_{F2}$, see \cite{Baranger1985,Mortensen1999}. The vertical line symbolizes the interface. 
At low temperatures and no applied bias voltage only electrons near the Fermi surfaces, indicated by the two circles, take part in transport processes. 
Flow from left to right requires $k_z > 0$ for electrons of the left-hand sphere.
When the size of the Fermi surfaces differs like in the figure, only electrons in the highlighted region can travel through the contact and find states in the highlighted region of the smaller right-hand sphere. 
The others are normal reflected.
In opposite direction,  normal reflection does not occur since all electrons from states of the left-hand  sphere with negative $k_z$ find states on the left-hand side within an angle $\Theta_c$ around the negative $k_z$ axis. 
The critical angle is $\Theta_c = \arcsin{(k_{F2}/k_{F1})}$ to satisfy conservation of parallel momentum $k_p$.}
  \label{ballistic pc} 
  \end{figure}

So far we have discussed normal and Andreev reflection in a one-dimensional model.
Both become more complicated in three dimensions as shown schematically in Figure \ref{ballistic pc}.
Since the momentum component parallel to the contact plane is assumed to be conserved but not the perpendicular one, only particles with angle of incidence  $\Theta \le \Theta_c = \arcsin{(k_{F2}/k_{F1})}$ can be transmitted, all others are reflected \cite{Baranger1985,Kashiwaya1996,Mortensen1999}. 
Thus, when the superconductor has the larger one of the wave numbers, only part of its Fermi surface can be probed by Andreev-reflection spectroscopy.
Total reflection does not occur in one dimension.
A dielectric tunnelling barrier appears stronger in three than in one dimension because electrons or holes with off-axis incidence have a longer path through the barrier.
Thus the one-dimensional BTK model over-estimates the $Z_b$-parameter \cite{Kashiwaya1996,Mortensen1999}. 
One hopes that the transition between one- and three-dimensional modelling would change the BTK parameters only slightly, and that the one-dimensional BTK model would remain valid and a good description of the experiments \cite{Kashiwaya1996,Mortensen1999,Daghero2010}.

In the one-dimensional BTK model $Z$ is often treated as a simple fit parameter without further consideration.
Using Andreev-reflection spectroscopy to determine material properties like the spin polarization or the symmetry of the superconducting order parameter requires  understanding normal reflection because it is essential for electrical transport across an interface.
We show here for contacts between superconducting indium (In) and several non-magnetic normal-conducting metals that in most cases the $Z$  parameter is probably neither related to a dielectric barrier nor to Fermi surface mismatch. 
Assuming that the contacts down to atomic size are in the diffusive regime would naturally explain our results.

\section{Experiments and results}

Point-contact experiments with superconducting indium (In) have a long history - junctions between In and normal metals have been investigated by Chien and Farrel \cite{Chien1975} even before the BTK theory became established.
Our contacts were fabricated using the shear (crossed wire) method by gently touching one sample wire with the other one as described by J. I. Pankove \cite{Pankove1966} and more recently by Chubov {\it et al.} \cite{Chubov1982}.
The In wires had $1.5\,$mm diameter to provide extra mechanical rigidity as much thinner wires would bend too easily when the contacts are made.
The  silver (Ag), palladium (Pd), platinum (Pt), and zinc (Zn) wires had $0.25\,$mm diameter. 
The contacts were measured below the critical temperature $T_c = 3.4\,$K of In down to 0.1\,K in the vacuum region of a dilution refrigerator.
A DC current $I$ with a small superposed AC component $dI$ is injected into 
the contact and the voltage drop $V+dV$ across the contact measured to obtain the $I(V)$ characteristics as well as the differential resistance spectrum $dV/dI(V)$.

Point contacts with In were more difficult to fabricate than those with aluminium (Al) \cite{Gloos2012}. 
Very often, when we tried to set the resistance, the contact either opened with a vacuum gap between the electrodes or closed with an extremely small resistance of order $1\,$m$\Omega$ which is unsuitable for spectroscopy. 
We attribute this behaviour to the softness of In.

\begin{figure}
\includegraphics[width=12cm]{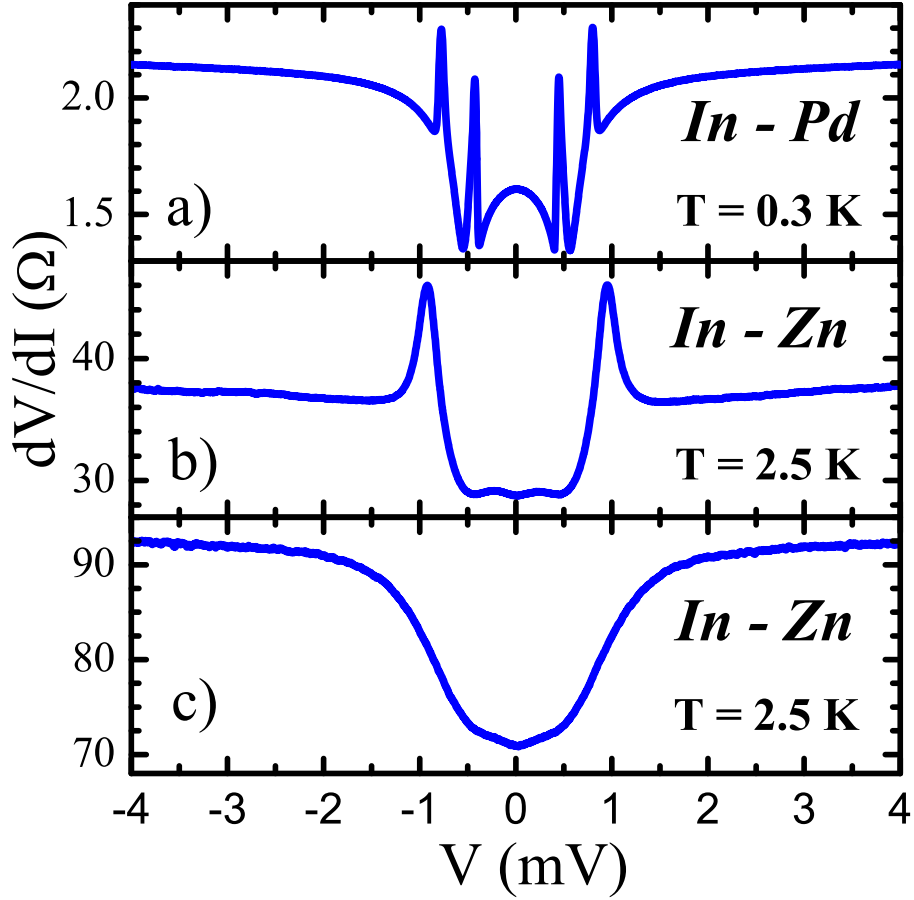}
\caption{(Color online) Differential resistance spectra $dV/dI$ versus bias voltage $V$ of contacts with In. Normal metal counter electrode and temperature are indicated. 
Spectrum a) shows an Andreev-reflection double-minimum structure. 
The two pairs of sharp side peaks make the analysis difficult.
Spectrum b) has one pair of side peaks and is slightly structured around zero bias.
Spectrum c) has no side peaks but a minimum at zero bias which appears to consist of two separate minima.
The shallow zero-bias minima in b) and c) could result from the proximity effect, while the side peaks in a) and b) stem from the self-magnetic field exceeding  a critical value in the contact region \cite{Gloos2009b}.}
  \label{spectra-In-BAD} 
  \end{figure}

We classify the spectra as follows:  81\% of the contacts had the typical Andreev reflection double-minimum anomaly and were further analysed.
5\% of the contacts had additional anomalies, like a dip at zero bias, that we tend to attribute to proximity-induced superconductivity in the normal metal. 
The remaining 14\% of the contacts showed spectra as in Figure \ref{spectra-In-BAD} with excessive side peaks or with anomalies that we do not really understand and which we do not consider further.
Table \ref{distribution} lists the details for the investigated normal conductors.

\begin{table}[position specifier]
\begin{center}
  \begin{tabular}{ c | c || c | c | c }
    \hline
    normal metal & total contacts & Andreev &  proximity-like & undefined \\  
    \hline \hline
    Ag & 83 & 71 & 2 & 10 \\ 
    \hline \hline
    Pd & 84 & 59 & 8 & 17 \\ 
    \hline \hline
    Pt & 26 & 26 & 0 & 0  \\ 
    \hline \hline
    Zn (at 2.5\,K) & 44 & 37 & 2 & 5 \\ \hline
    \hline
  \end{tabular}
\end{center}\caption{Distribution of contact type: 'Andreev' denotes contacts that could be analysed, 'proximity-like' contacts look like those where superconductivity has been induced in the normal metal, and 'undefined' are all others which can not be clearly identified.}
\label{distribution}
\end{table}

\begin{figure}
\includegraphics[width=12cm]{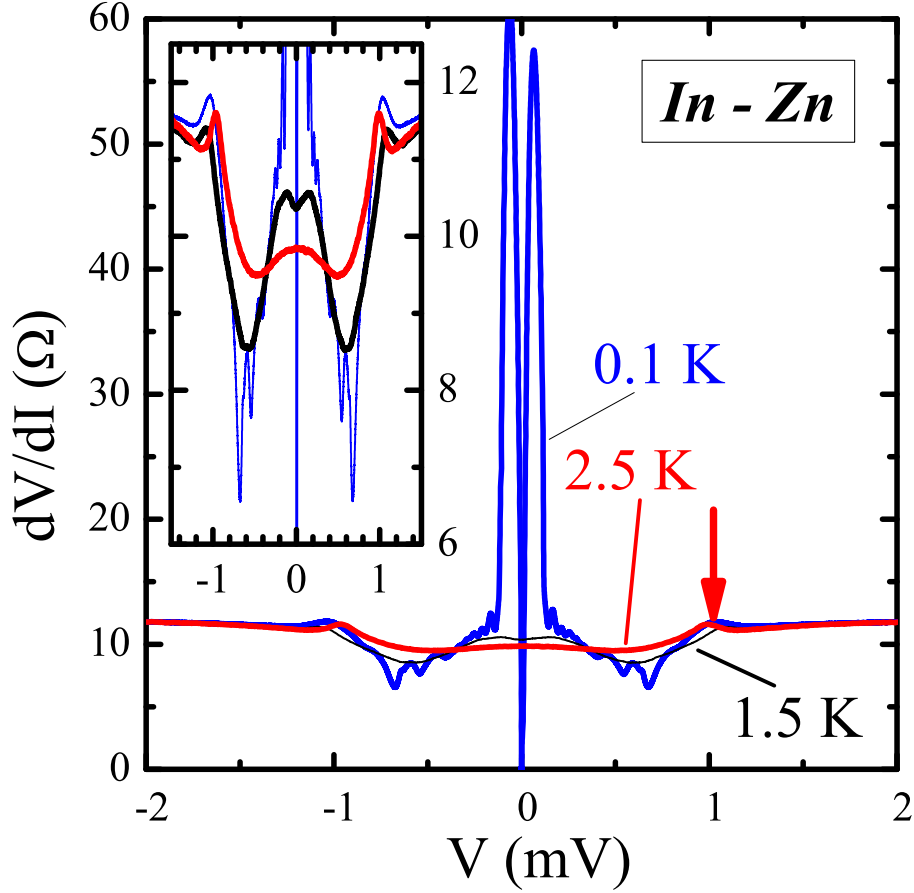}
\caption{(Color online) Differential resistance spectra $dV/dI$ versus bias voltage $V$ of an In - Zn junction at the indicated temperatures. 
At $T = 0.1\,$K the Josephson-type anomaly at zero bias as well as multiple Andreev reflection are clearly visible. 
At $T = 1.5\,$K the little dip at zero bias could indicate proximity-induced superconductivity in Zn. 
It is completely suppressed at $T = 2.5\,$K. 
The spectra of this contact show a side peak (arrow), which moves slightly with temperature and is possibly caused by the self-magnetic field.
The inset displays the details of the differential resistance around zero bias with emphasis on the spectra at 1.5\,K and 2.5\,K, respectively.}
  \label{spectra-In-Zn-PE} 
  \end{figure}

At low temperatures  the contacts with Zn had typical Josephson-type characteristics with multiple Andreev reflection  as presented in Figure \ref{spectra-In-Zn-PE}.
The temperature has to be raised to 2.5\,K, well above the critical temperature $T_c = 0.87\,$K, of Zn to suppress the Josephson-type and proximity-like anomalies in most junctions.
Although this procedure strongly reduces the magnitude of the Andreev-reflection signal, the $Z$-parameter can still be extracted.

The chosen spectra were analysed using a modified BTK theory that includes Dynes' lifetime parameter $\Gamma$ \cite{Plecenik1994}, so that in total the model contains three adjustable parameters.
The normal resistance was defined as the differential resistance at large bias voltages.
Side peaks at finite bias voltage, for example due to the self-magnetic field \cite{Gloos2009b}, were usually easy to recognize and did therefore not affect the analysis with respect to the $Z$-parameter. 

\begin{figure}
\includegraphics[width=12cm]{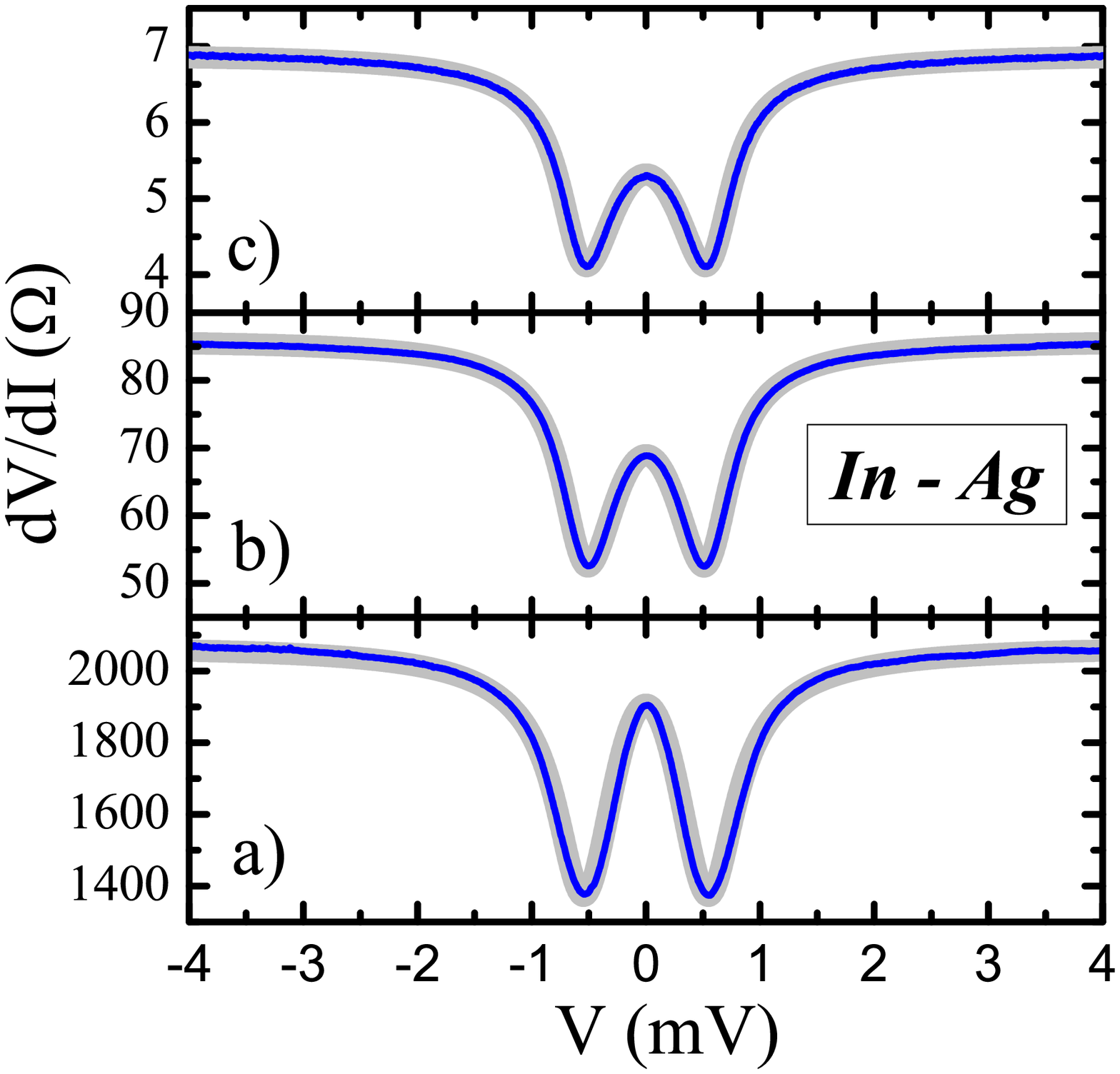}
\caption{(Color online) Selected differential resistance spectra $dV/dI$ versus bias voltage $V$ of In - Ag contacts at $T = 0.7\,K$ (thin lines). The underlying grey curves are fits with the modified BTK model \cite{Plecenik1994}. Fit parameters are in Table \ref{fitparameters}.}
  \label{spectra-In-Ag} 
  \end{figure}

\begin{figure}
\includegraphics[width=12cm]{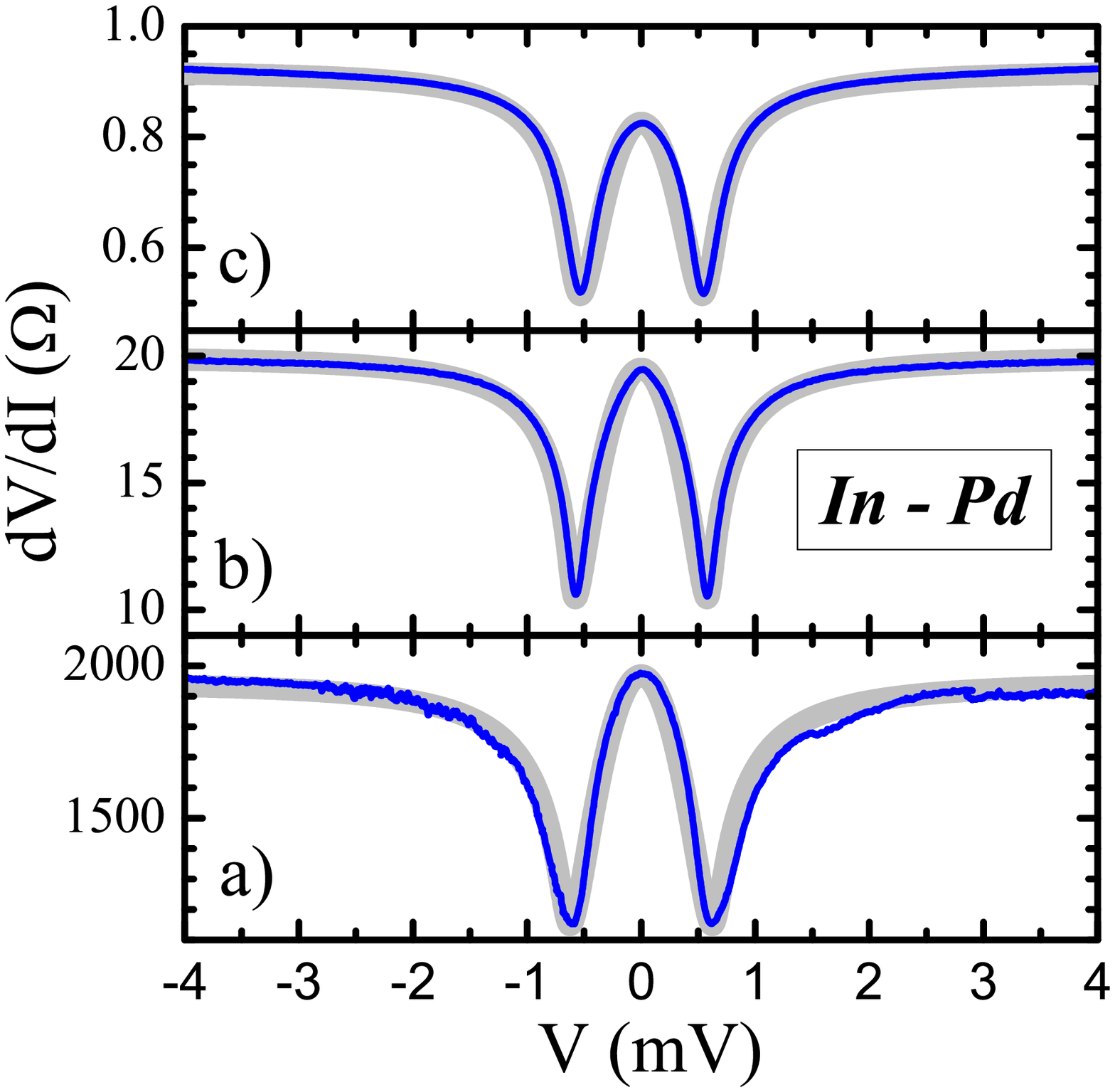}
\caption{(Color online) Selected differential resistance spectra $dV/dI$ versus bias voltage $V$ of In - Pd contacts at low temperatures (thin lines). The underlying grey curves are fits with the modified BTK model \cite{Plecenik1994}. Temperatures and fit parameters are in Table \ref{fitparameters}.}
  \label{spectra-In-Pd} 
  \end{figure}

\begin{figure}
\includegraphics[width=12cm]{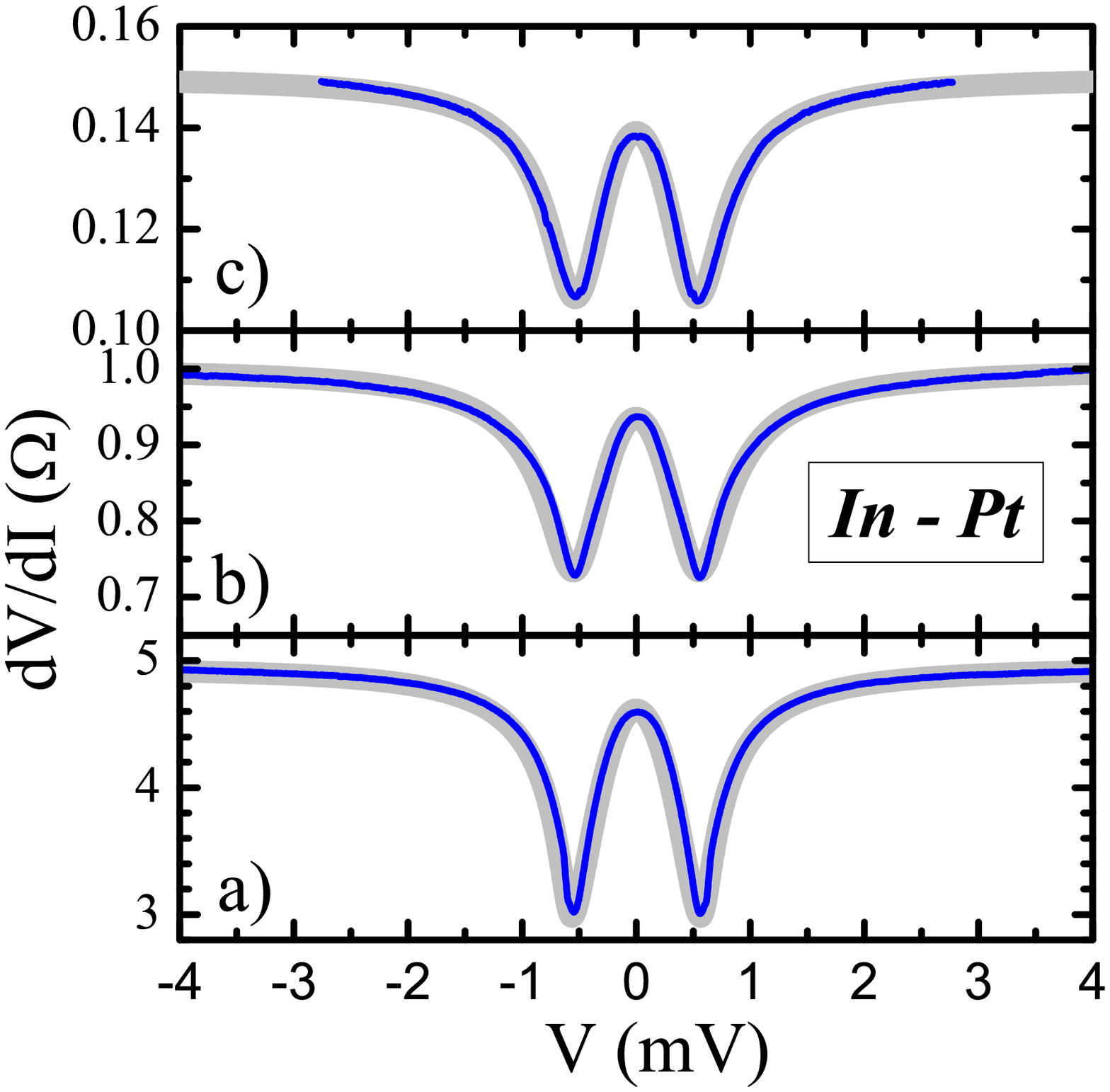}
\caption{(Color online) Selected differential resistance spectra $dV/dI$ versus bias voltage $V$ of In - Pt contacts at low temperatures (thin lines). The underlying grey curves are fits with the modified BTK model \cite{Plecenik1994}. Temperatures and fit parameters are in Table \ref{fitparameters}.}
  \label{spectra-In-Pt} 
  \end{figure}

\begin{figure}
\includegraphics[width=12cm]{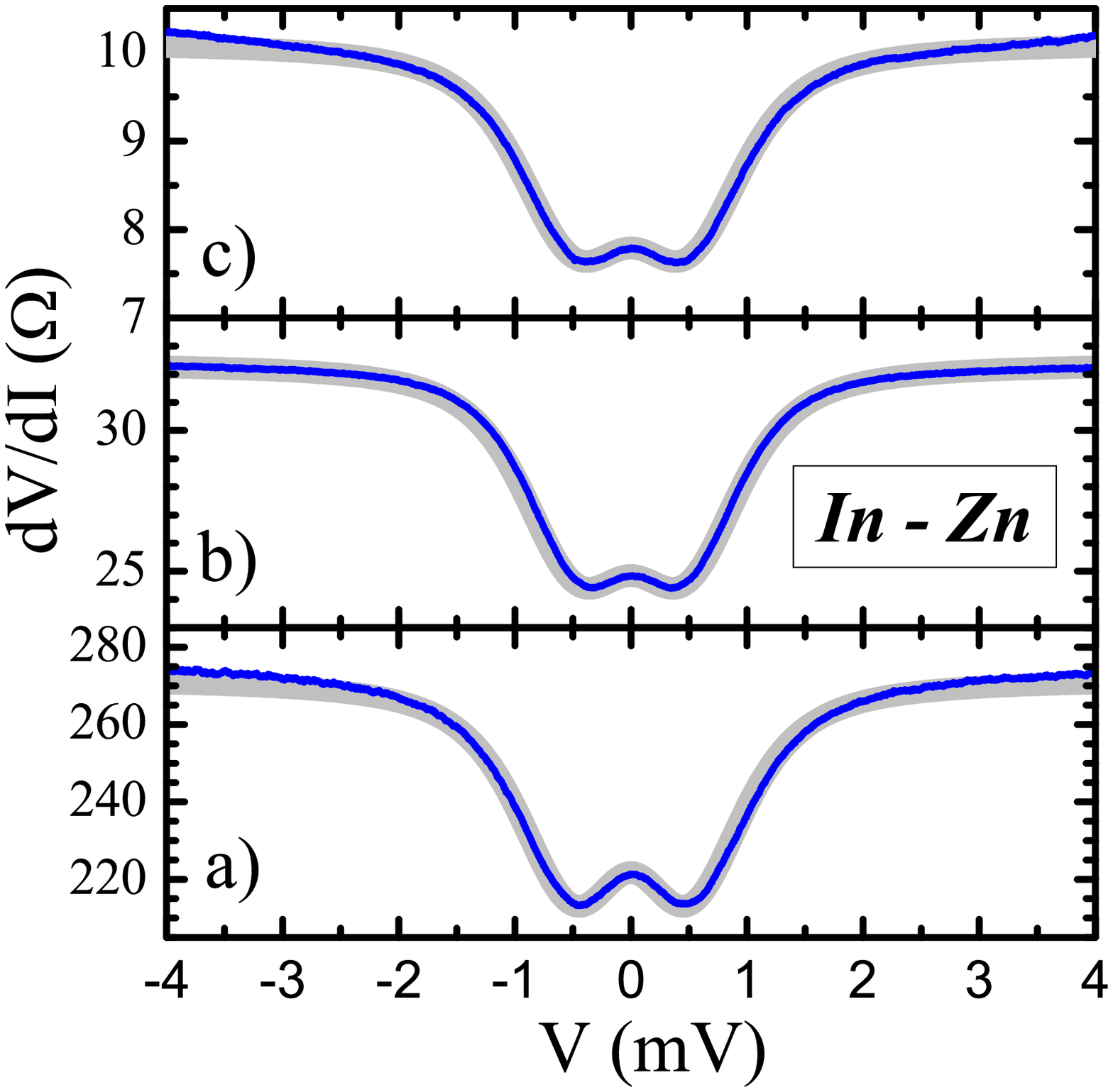}
\caption{(Color online) Selected differential resistance spectra $dV/dI$ versus bias voltage $V$ of In - Zn contacts at $T = 2.5\,K$ (thin lines). The underlying grey curves are fits with the modified BTK model \cite{Plecenik1994}. Fit parameters are in Table \ref{fitparameters}.}
  \label{spectra-In-Zn} 
  \end{figure}

Figures \ref{spectra-In-Ag}, \ref{spectra-In-Pd}, \ref{spectra-In-Pt}, and \ref{spectra-In-Zn}
show selected spectra of superconducting In in contact with Ag, Pd, Pt, and Zn over the accessed resistance range together with a fit using the modified BTK model. 
Note that this is a one-dimensional model valid for ballistic transport.
Table \ref{fitparameters} summarizes the extracted fit parameters.

\begin{table}[position specifier]
\begin{center}
  \begin{tabular}{ c || c | c || c | c | c }
    \hline
    metal & $R (\Omega)$ & $T({\text{K}})$ &  $2\Delta_0$ (meV) & $\Gamma$ (meV) & $Z$ \\  
    \hline \hline
       & 6.9   & 0.7 & 1.18 & 0.025 & 0.420 \\ 
    Ag & 86.0  & 0.7 & 1.14 & 0.030 & 0.445 \\ 
       & 2071 & 0.7 & 1.17 & 0.063 & 0.535 \\ 
    \hline \hline
       & 0.92 & 0.40 & 1.14 & 0.020 & 0.505 \\ 
    Pd & 20.0 & 0.13 & 1.20 & 0.025 & 0.550 \\ 
       & 1950 & 0.1 & 1.30 & 0.060 & 0.580 \\ 
    \hline \hline
       & 0.15 & 0.5 & 1.14 & 0.110 & 0.540 \\ 
    Pt & 1.00 & 0.3 & 1.17 & 0.145 & 0.545 \\ 
       & 4.95 & 0.1 & 1.19 & 0.060 & 0.530 \\ 
    \hline \hline
       & 10.16 & 2.5 & 1.30 & 0.045 & 0.485 \\
    Zn & 32.4 & 2.5 & 1.22 & 0.030 & 0.495 \\ 
       & 275 & 2.5 & 1.23 & 0.040 & 0.555 \\ 
    \hline
  \end{tabular}
\end{center}\caption{Normal contact resistance, measurement temperature, and BTK parameters of the spectra shown in Figures \ref{spectra-In-Ag}, \ref{spectra-In-Pd}, \ref{spectra-In-Pt}, and \ref{spectra-In-Zn}.}
\label{fitparameters}
\end{table}

\begin{figure}
\includegraphics[width=12cm]{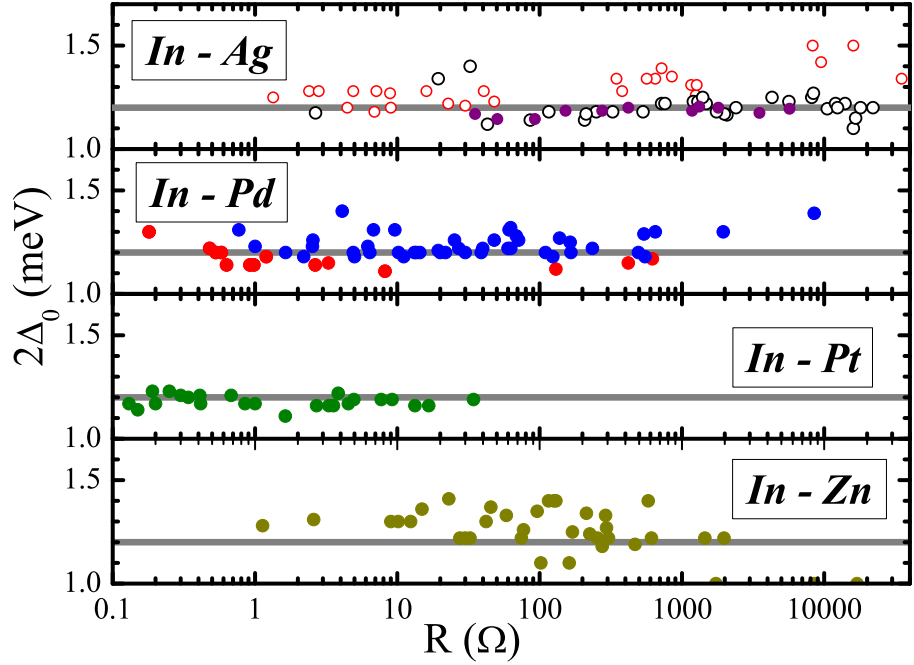}
\caption{(Color online) Superconducting energy gap $2\Delta_0 = 2\Delta(T \rightarrow 0)$ of In extracted from the point-contact spectra using the modified BTK theory \cite{Plecenik1994} versus normal state resistance $R$.
Different symbols mark separate measurement series. 
For In - Ag contacts we have two measurement series at $T = 0.7\,$K (open symbols) and one at $T = 0.1\,$K. 
All In - Zn contacts were measured at $T = 2.5\,$K.
The In - Pd and the In - Pt contacts were measured at low temperature down to $T = 0.1\,$K.
The solid lines are $2\Delta_0 = 1.20\,$meV as a guide to the eye.}
  \label{para-2d0} 
  \end{figure}
\begin{figure}
\includegraphics[width=12cm]{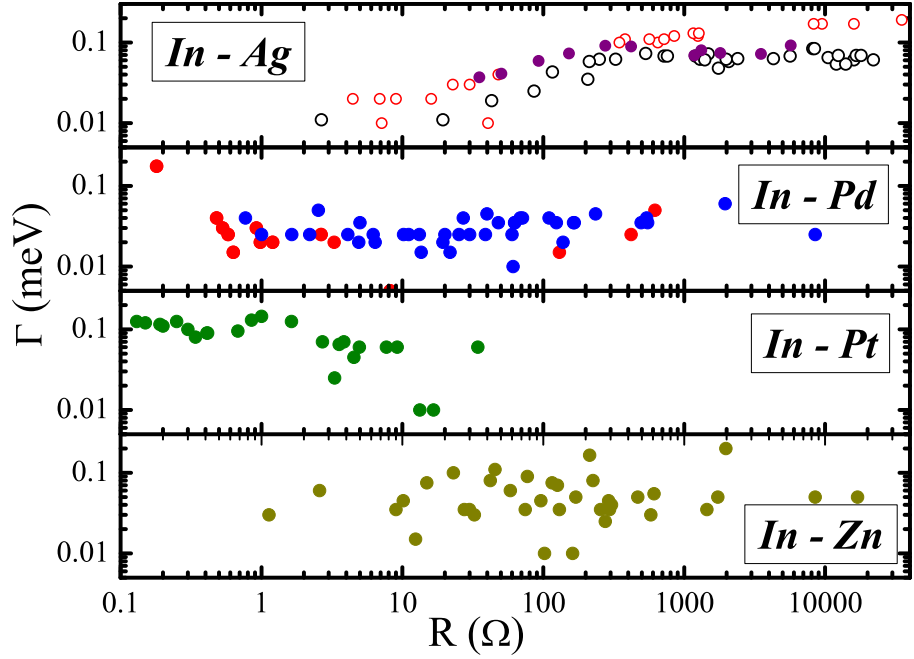}
\caption{(Color online) Dynes lifetime parameter $\Gamma$ of In in contact with the indicated normal metals extracted from the point-contact spectra using the modified BTK theory \cite{Plecenik1994} versus normal state resistance $R$.
Different symbols mark separate measurement series. 
For In - Ag contacts we have two measurement series at $T = 0.7\,$K (open symbols) and one at $T = 0.1\,$K. 
All In - Zn contacts were measured at $T = 2.5\,$K.
The In - Pd and the In - Pt contacts were measured at low temperature down to $T = 0.1\,$K.
Note the logarithmic $\Gamma$-scale.}
  \label{para-g} 
  \end{figure}
\begin{figure}
\includegraphics[width=12cm]{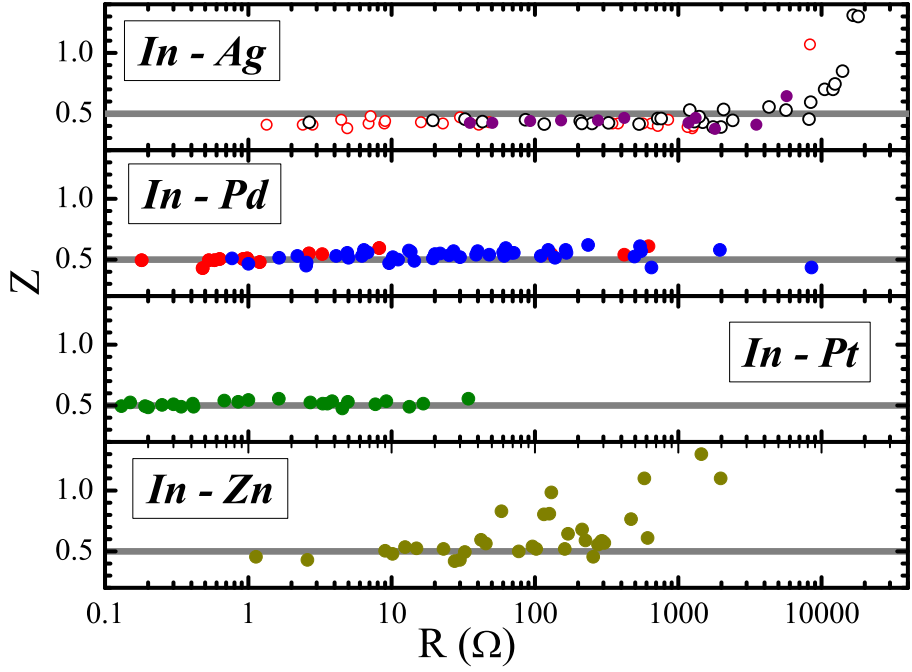}
\caption{(Color online) $Z$ parameter of normal reflection of contacts between In and the indicated normal conductors extracted from the point-contact spectra using the modified BTK theory \cite{Plecenik1994} versus normal state resistance $R$.
Different symbols mark separate measurement series. 
For In - Ag contacts we have two measurement series at $T = 0.7\,$K (open symbols) and one at $T = 0.1\,$K. 
All In - Zn contacts were measured at $T = 2.5\,$K.
The In - Pd and the In - Pt contacts were measured at low temperature down to $T = 0.1\,$K.
The solid lines are  $Z=0.5$ as a guide to the eye.}
  \label{para-z} 
  \end{figure}

Figures \ref{para-2d0}, \ref{para-g}, and \ref{para-z}   show the derived parameters $2\Delta_0 = 2\Delta(T \rightarrow 0)$, $\Gamma$, and $Z$ as function of normal resistance $R$ for contacts between superconducting In and normal conducting Ag, Pd, Pt, and Zn, respectively. 
The energy gap $2\Delta_0 \approx 1.2\,$meV is roughly constant from $\sim 0.1\,\Omega$ up to $\sim 10\,$k$\Omega$.
Most contacts have a $\Gamma$ that varies between $10\,\mu$eV and $100\,\mu$eV without a clear tendency. 
The lifetime parameter $\Gamma$ grows slightly with increasing resistance only for the contacts with Ag, independently of the temperature. 
The $Z$-parameter stays constant at 0.5 from $0.1\,\Omega$ up to several $1\,$k$\Omega$ and varies by less than $\pm 0.1$ from one normal conductor to another one.
Larger $Z$-values appear for In - Ag junctions in the $10\,$k$\Omega$ range and for the In - Zn contacts above $\sim 100\,\Omega$.

\section{Discussion}

The BTK parameters of our contacts with superconducting In correspond  well with those of superconducting Al \cite{Gloos2012}.
Unlike $2\Delta_0(R)$ of Al, that increases with $R$, the In contacts have a rather constant $2\Delta_0 \approx 1.2\,$meV.
A systematic increase of $\Gamma$ with $R$ like for the Al contacts has only been found for In - Ag contacts, but to a lesser degree. 
For contacts with the other normal-conducting metals $\Gamma$ ranges from $10\,\mu$eV to $100\,\mu$eV.
However, most notable is $Z\approx 0.5$ over up to five orders of magnitude in normal state resistance $R$ like for superconducting Al \cite{Gloos2012}.
Point contacts with superconducting Nb, measured at $T = 4.2\,$K, have slightly larger average Z but with a wider variation \cite{Tuuli2011}.
In the discussion we will focus on the seemingly universal value of $Z$.

{\bf Contact diameter.}
We estimate the contact diameter $d$ with the ballistic Sharvin resistance $R = 8 R_K / (dk_F)^2$ where $R_K = h/e^2$.
In free-electron approximation the used metals have Fermi wave numbers $k_F \approx 14\,$nm$^{-1}$ \cite{Ashcroft1976}. 
Then a $1\,\Omega$ contact has a diameter of $d \approx 32\,$nm, assuming circular symmetry, or $\sim \! 830\,{\text{nm}}^2$ cross-sectional area.
Thus our study covers contact areas from $10000\,$nm$^2$ to less than $1\,$nm$^2$. 
According to the residual normal-state resistivity of the bulk metals we estimate elastic mean free paths of $l_0 \gtrsim 10\,$nm.
Therefore contacts with normal resistance $R \gg 1\,\Omega$ should be ballistic, justifying our use of the BTK model.

{\bf Superconducting energy gap.}
The energy gap $2\Delta_0 \approx 1.2\,$meV is almost constant from around $0.1\,\Omega$ up to about $10\,$k$\Omega$.
The ratio $2\Delta_0/k_B T_c \approx 4.05$ is larger than the BCS value 3.52 and the tunnelling-derived bulk value of 3.58 \cite{Averill1972}.
A slightly enhanced energy gap has been found earlier for In break junctions \cite{Gloos1999}.
These deviations could be caused by the pressure or lattice distortion at the contact.

{\bf Lifetime parameter.}
The lifetime parameter $\Gamma$ was originally introduced by Dynes {\it et al.} \cite{Dynes1978} to describe the enhanced pair breaking in superconducting lead alloy films.
A point contact could cause pair breaking since it disturbs locally the crystal lattice symmetry.
As an alternative explanation, Raychaudhuri {\it et al.} \cite{Raychaudhuri2004} have suggested that an inhomogeneous superconducting gap in the contact region could also lead to a finite $\Gamma$-value.
A similar approach was used by Bobrov {\it et al.} \cite{Bobrov2005}.
The order parameter can be reduced at the interface, and Cooper pairs can leak into the normal metal.
This could also explain the zero-bias dip in the spectra of In - Zn junctions well above the critical temperature of Zn as shown in Figure \ref{spectra-In-Zn-PE}.

{\bf Dielectric barrier.} 
Metal surfaces usually oxidize when they are exposed to air.
Because our setup does not allow transferring the samples to the refrigerator under ultra-high vacuum conditions, the sample surfaces are very likely oxidized.
For example, a typical metal oxide Al$_2$O$_3$ on bulk Al has a thickness of $w \approx 1\,$nm and a potential height of $\Phi \approx 2\,$eV  \cite{Gloos2003}. 
Assuming a Fermi velocity $v_F = 1500\,$km/s \cite{Ashcroft1976} we obtain $Z = \Phi w / \hbar v_F \approx 2$, varying from 1 to more than 10.
The $Z \approx 0.5$ observed in our experiments corresponds to a significantly weaker tunnel barrier.
We have found considerably larger $Z$-values occasionally.

Only when the contacts are very small, the intrinsic cleaning process of the shear method, when the two sample wires slide along each other before the contact forms, might fail and preserve a nearly undisturbed dielectric layer of the atomic-size contacts. 
The $Z$-value increases in the k$\Omega$-range towards the transition to vacuum tunneling.
Such high-resistance contacts consist of a few conduction channels, each with its own transmission coefficient. 
For the Zn contacts the deviations appear already at $100\,\Omega$, possibly indicating that ZnO is more difficult to remove or to damage than the other metal oxides.
Or it could be due to ZnO being a semiconductor instead of an insulator \cite{ZnO}.

The argument against a tunnelling barrier is the independence of the $Z$-parameter from contact size and  normal metal electrode.
In addition, reflection at a dielectric barrier should lead to a strong variation of $Z$, depending on how a specific contact is made, because the transmission probability depends exponentially on the barrier width and height \cite{Simmons1964}.
Therefore we would have expected $Z$ not to converge to a single value, but to vary from almost zero, the lower bound defined by Fermi surface mismatch according to Eq. \ref{Z parameter}, to $Z \gg 10$ with a thick and nearly undisturbed oxide layer.

{\bf Fermi surface mismatch.} 
Electrons as well as Andreev-reflected holes travel through a dielectric tunneling barrier with a certain probability, while the rest are normal reflected.
However, normal reflection due to Fermi surface mismatch is different.
This becomes obvious in the three-dimensional model: using the notation of Figure \ref{ballistic pc}, electrons with the direction of incidence $\Theta > \Theta_c = \arcsin{(k_{F2}/k_{F1})}$ can not cross the interface \cite{Baranger1985,Kashiwaya1996,Mortensen1999}, and therefore can not take part in Andreev reflection.
Since only electrons that have been transmitted can be Andreev-reflected, the retro-reflected holes have already the right properties to be transmitted back through the interface and are treated as described by the BTK model. 
This isotropic case might approximate contacts between metals with nearly spherical Fermi surface, like potassium (K) or Ag,  but it should fail for the transition metals, like Nb or tantalum (Ta).
Their Fermi surfaces, that determine the transport processes, can consist of multiple sheets or isolated pockets \cite{Choy2000}.
Therefore one should expect a pronounced difference with respect to Andreev and normal reflection when nearly free-electron metals, both normal and superconducting, are compared to the transition metals.
This has not been found: for example the shape of the Andreev reflection spectra of contacts with Nb looks very similar to that of spectra with Al \cite{Gloos2012b}. 

A practical argument against Fermi surface mismatch in our point contacts is, like discussed above with respect to a dielectric barrier, the small variation of the $Z$-parameter. 
Each time we make a new contact, the orientation of the crystallites that form it also changes, and the $Z$-parameter should change accordingly.
The same should happen when the normal conductor is replaced.
This is not observed.

\begin{figure}
\includegraphics[width=12cm]{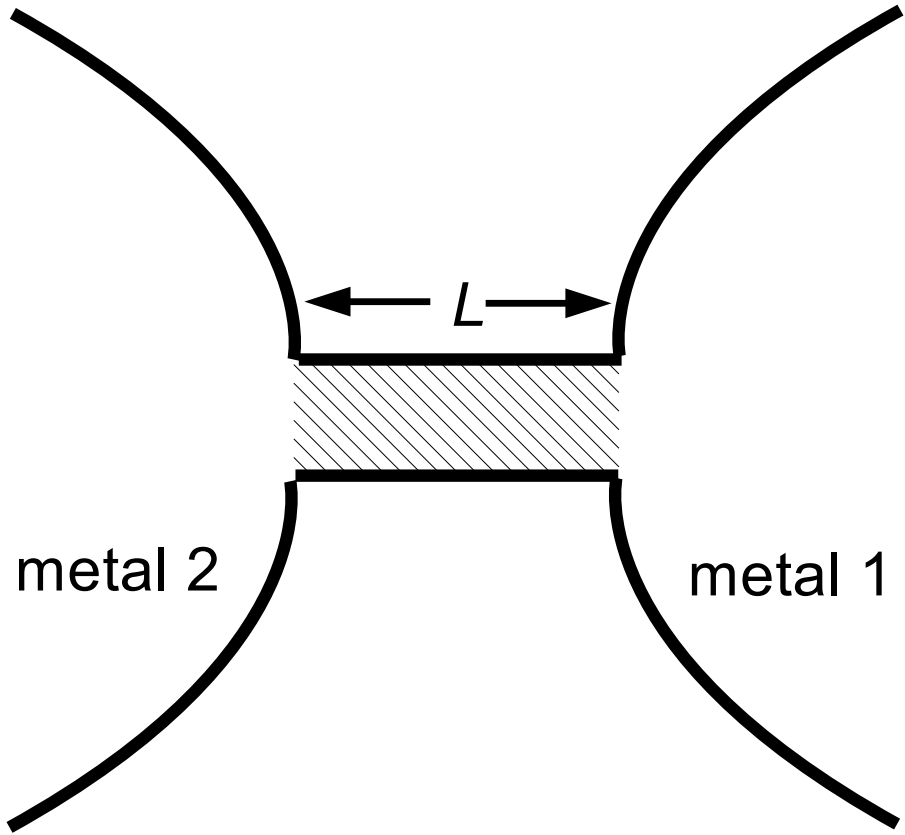}
\caption{(Color online) Schematics of a diffusive point contact between two different metals.
The contact (hatched area) has a spatial extention $L \gg l_0$.
Its elastic electron mean free path $l_0$ is not necessarily the same as in the bulk electrodes.}
  \label{diffusive pc} 
  \end{figure}

{\bf Diffusive limit.} 
Figure \ref{diffusive pc} shows schematically a diffusive contact between two metal electrodes.
The size of the link $L$ between the electrodes corresponds to the point-contact diameter $d$.
The contact is diffusive when this length is much larger than the elastic electron mean path but shorter than the diffusion length $l_0 \ll  L \le \Lambda$. 
Under these conditions electrons flowing through the constriction suffer many elastic scattering processes, loosing the directional information of their momentum, but keep their energy at zero bias voltage or gain the excess energy $eV$ when a bias voltage is applied.
This is valid even if the electrodes have a much larger electron mean free path or superconducting coherence length.

Naidyuk {\it et al.} \cite{Naidyuk1996b} have already noticed that a $Z$-parameter between 0.4 and 0.5 is often found for superconducting point contacts, including those with heavy-fermion and high-temperature superconductors.
Naidyuk and Yanson \cite{Naidyuk2005} have suggested that the contacts could be in the diffusive limit as described by Mazin {\it et al.} \cite{Mazin2001} who showed that the spectra of diffusive contacts without an interface barrier are almost identical to those of ballistic contacts with a finite $Z$-parameter close to 0.55.
Artemenko {\it et al.} \cite{Artemenko1979} had derived this result earlier, and it is also mentioned in the seminal paper by Blonder {\it et al.} \cite{Blonder1982}.
Note  that typical contacts used for Andreev-reflection spectroscopy usually have a resistance of less than $100\,\Omega$ and, thus, a diameter of more than a few nm.
Such contacts could easily be in the diffusive regime.

According to our earlier results on In break junctions \cite{Gloos2009b} with an elastic mean free path of 25\,nm in the contact region, our contacts with normal resistance below $1\,\Omega$ are probably in the diffusive regime.
When the contacts are made smaller, they should become ballistic.
However, we can not notice any change of contact properties that would indicate a transition between the two transport regimes.
Therefore, in our earlier paper on contacts with superconducting Al \cite{Gloos2012} we have implicitly assumed that large resistance contacts would be ballistic, and, without noticing a changing behaviour, that they would stay ballistic when the contact resistance is reduced.
Maybe the converse argument is more appropriate.

\section{Conclusion}
Understanding the role of normal reflection is an especially important topic for applying Andreev-reflection spectroscopy to investigate unconventional superconductors or the local spin polarization of ferromagnets. 
We have found that the BTK $Z$-parameter of interfaces with superconducting In does neither depend sensitively on the size of the contacts nor on the normal-conducting counter electrode.
This agrees  well with our earlier data for interfaces with superconducting Al.
The tiny variation of $Z$ over a wide range of contact areas has lead us to conclude that a dielectric tunneling barrier as well as Fermi surface mismatch do not contribute significantly.
One explanation for this behaviour could be that all of our contacts are in the diffusive limit.
Our results question the nature of the parameters derived from one-dimensional ballistic models in point-contact Andreev-reflection spectroscopy and call for an investigation of the effects of dimensionality and transport regime.

\begin{acknowledgments}
E. T.  acknowledges a two-year grant from the Graduate School of Materials Research (GSMR), 20014 Turku, Finland. We thank Yu. G. Naidyuk for discussions and the Jenny and Antti Wihuri Foundation for financial support.
\end{acknowledgments}


\newpage 

%
%

\end{document}